\documentclass[12pt]{article}
  \usepackage{amsfonts}
\usepackage{amsmath}

\title{Progress in  Gauge  Invariant  Lagrangian Construction for Massive Higher
Spin Fields}
\author{I.L. Buchbinder\footnote{joseph@tspu.edu.ru},
 V.A. Krykhtin\footnote{krykhtin@mph.phtd.tpu.edu.ru}}
\date{\it Department of Theoretical Physics Tomsk State Pedagogical
University \\ Tomsk, 634041, Russia}

\begin{document}
\maketitle
\begin{abstract}
We review the recently developed general gauge invariant approach to
Lagrangian construction for massive higher spin fields in Minkowski
and AdS spaces of arbitrary dimension. Higher spin Lagrangian,
describing the dynamics of the fields with any spin, is formulated
with help of BRST-BFV operator in auxiliary Fock space. No off-shell
constraints on the fields and gauge parameters are imposed. The
construction is also applied to tensor higher spin fields with index
symmetry corresponding to a multirow Young tableau.
\end{abstract}

\section{Introduction}
Higher spin filed problem attracts much attention during a long
time. At present, there exist the various approaches to this problem
although the many aspects are still far to be completely clarified
(see e.g. \cite{reviews} for recent reviews of massless higher spin
field theory). This paper is a brief survey of recent state of gauge
invariant approach to massive higher spin field theory.

The standard BFV or BRST-BFV construction (see the reviews
\cite{BRST}) arose at operator quantization of dynamical systems
with first class constraints. The systems under consideration are
characterized by first class constraints in phase space $T_a$,
$[T_a,T_b]=f^c_{ab}T_c$. Then BRST-BFV charge is introduced
according to the rule
\begin{eqnarray}
Q&=&\eta^aT_a+\frac{1}{2}\eta^b\eta^af^c_{ab}{\cal{}P}_c,
\qquad
\qquad
\qquad
Q^2=0,
\label{eq1}
\end{eqnarray}
where $\eta^a$ and ${\cal{}P}_a$ are canonically conjugate ghost
variables (we consider here the case $gh(T)=0$, then $gh(\eta^a)=1$,
$gh({\cal{}P}_a)=-1$) satisfying the relations
$\{\eta^a,{\cal{}P}_b\}=\delta^a_b$. After quantization the BRST-BFV
charge becomes an Hermitian operator acting in extended space of
states including ghost operators, the physical states in the
extended space are defined by the equation $Q|\Psi\rangle=0$. Due to
the nilpotency of the BRST-BFV operator, $Q^2=0$, the physical
states are defined up to transformation $|\Psi'\rangle =
|\Psi\rangle+Q|\Lambda\rangle$ which is treated  as a gauge
transformation.

Application of BRST-BFV construction in the higher spin field theory
is inverse to above quantization problem. The initial point are
equations, defying the irreducible representations of Poincare or
AdS groups with definite spin and mass, the BRST-BFV operator is
constructed on the base of these constraints and finally the higher
spin Lagrangian is found on the base of BRST-BFV operator. Generic
procedure looks as follows. The equations defining the
representations are treated as the operators of first class
constraints in some auxiliary Fock space. However, in the higher
spin field theory the part of these constraints are non-Hermitian
operators and in order to construct a Hermitian BRST-BFV operator we
have to involve the operators which are Hermitian conjugate to the
initial constraints and which are not the constraints. Then for
closing the algebra to the complete set of operators we must add
some more operators which are not constraints as well. Because of
presence of such operators the standard BRST-BFV construction can
not be applied it its literal form. However, as we will see, this
problem can be solved.


\section{Massive bosonic field}

We illustrate the method used for Lagrangian construction on the
base of massive bosonic field in Minkowski $d$-dimensional space. It
is well known that the totally symmetric tensor field
$\Phi_{\mu_1\cdots\mu_s}$, describing the irreducible spin-$s$
massive representation of the Poincare group must satisfy the
following constraints
\begin{eqnarray}
&
(\partial^2+m^2)\Phi_{\mu_1\cdots\mu_s}=0,
\qquad
\partial^{\mu_1}\Phi_{\mu_1\mu_2\cdots\mu_s}=0,
\qquad
\eta^{\mu_1\mu_2}\Phi_{\mu_1\cdots\mu_s}=0.
\label{irrep0}
\end{eqnarray}

In order to describe all higher integer spin fields
simultaneously it is convenient to introduce Fock space
generated by creation and annihilation operators $a_\mu^+$,
$a_\mu$ with vector Lorentz index $\mu=0,1,2,\ldots,d-1$
satisfying the commutation relations
\begin{eqnarray}
\bigl[a_\mu,a_\nu^+\bigr]=-\eta_{\mu\nu},
\qquad
\eta_{\mu\nu}=(+,-,\ldots,-).
\end{eqnarray}
Then we define the operators
\begin{eqnarray}
\label{l0}
&l_0=-p^2+m^2,
\qquad
l_1=a^\mu p_\mu,
\qquad
l_2=\frac{1}{2}a^\mu a_\mu,
\end{eqnarray}
where
$p_\mu=-i\frac{\partial}{\partial{}x^\mu}$.
These operators act on states in the
Fock space
\begin{eqnarray}
|\Phi\rangle
&=&
\sum_{s=0}^{\infty}\Phi_{\mu_1\cdots\mu_s}(x)
a^{\mu_1+}\cdots a^{\mu_s+}|0\rangle
\label{gstate}
\end{eqnarray}
which describe all integer spin fields simultaneously if
the following constraints on the states take place
\begin{eqnarray}
&
l_0|\Phi\rangle=0,
\qquad
l_1|\Phi\rangle=0,
\qquad
l_2|\Phi\rangle=0.
\label{01}
\end{eqnarray}
If constraints (\ref{01}) are fulfilled for the general state
(\ref{gstate}) then constraints (\ref{irrep0}) are fulfilled for
each component $\Phi_{\mu_1\cdots\mu_s}(x)$ in (\ref{gstate}) and
hence the relations (\ref{01}) describe all free massive higher spin
bosonic fields simultaneously. Our purpose is to describe the
Lagrangian construction for the massive higher spin fields on the
base of BRST-BFV approach, therefore first what we should find is
the Hermitian BRST-BFV operator. It means, we should have a system
of Hermitian constraints. In the case under consideration the
constraint $l_0$ is Hermitian,
 $l_0^+=l_0$, however the constraints $l_1,$ $l_2$ are not Hermitian.
We extend the set of the constraints $l_0,$ $l_1,$ $l_2$ adding two
new operators $l_1^+=a^{\mu+}p_\mu$, $l_2^+=\frac{1}{2}a^{\mu+}a_\mu^+$.
As a result, the set of operators
$l_0,$ $l_1,$ $l_2,$ $l_1^+,$ $l_2^+$
is invariant under Hermitian conjugation.
We want to point out that operators $l_1^+$, $l_2^+$ are not
constraints on the space of bra-vectors (\ref{gstate}) since
they may not annihilate the physical states. Taking Hermitian
conjugation of (\ref{01}) we see that $l_1^+$, $l_2^+$ (together
with $l_0$) are constraints on the space of
bra-vectors.

Algebra of the operators $l_0$, $l_1$, $l_1^+$, $l_2$, $l_2^+$ is
open in terms of commutators of these operators. We will suggest the
following procedure of consideration. We want to use the BRST-BFV
construction in the simplest (minimal) form corresponding to closed
algebras. To get such an algebra we add to the above set of
operators, all operators generated by the commutators of $l_0$,
$l_1$, $l_1^+$, $l_2$, $l_2^+$. Doing such a way  we obtain two new
operators\footnote{The resulting algebra can be found in
\cite{0505092}.}
\begin{eqnarray}
m^2
&\qquad\mbox{and}\qquad&
g_0=-a_\mu^+a^\mu+\frac{d}{2}.
\label{mg}
\end{eqnarray}

Let us emphasize once again that operators $l_1^+$, $l_2^+$ are not
constraints on the space of ket-vectors. The constraints in space of
ket-vectors are $l_0$, $l_1$, $l_2$ and they are the first class
constraints in this space. Analogously, the constraints in space of
bra-vectors are $l_0$, $l_1^+$, $l_2^+$ and they also are the first
class constraints but only in this space, not in space of
ket-vectors. Since the operator $m^2$ is obtained from the
commutator
\begin{eqnarray}
[l_1,l_1^+]=l_0-m^2,
\end{eqnarray}
where $l_1$ is a constraint in the space of ket-vectors and
$l_1^+$ is a constraint in the space of bra-vectors, then it can
not be regarded as a constraint neither in the ket-vector space
nor in the bra-vector space.
Analogously the operator $g_0$ is obtained from the commutator
\begin{eqnarray}
[l_2,l_2^+]=g_0,
\end{eqnarray}
where $l_2$ is a constraint in the space of ket-vectors and
$l_2^+$ is a constraint in the space of bra-vectors.
Therefore $g_0$ can not also be regarded as a constraint neither
in the ket-vector space nor in the bra-vector space.

One can show that a straightforward use of BRST-BFV construction as
if all the operators $l_0,$ $l_1,$ $l_2,$ $l_1^+,$ $l_2^+$, $g_0$,
$m^2$ are the first class constraints doesn't lead to the proper
equations (\ref{01}) for any spin. This happens because among the
above hermitian operators there are operators which are not
constraints ($g_0$ and $m^2$ in the case under consideration) and
they bring two more equations (in addition to (\ref{01})) onto the
physical field (\ref{gstate}). Thus we must somehow get rid of these
supplementary equations.

The method of avoiding the supplementary equations consists in
constructing the new enlarged expressions for the operators of the
algebra, so that the Hermitian operators which are not constraints
will be zero.

Let us act as follows.
We enlarge
the representation space of the operator algebra
by introducing the additional (new) creation and
annihilation operators
and enlarge expressions for the operators (see \cite{0505092}
for more details)
\begin{eqnarray*}
l_i&\longrightarrow&L_i=l_i+l_i',
\qquad
\qquad
l_i=\{l_0,l_1,l_1^+,l_2,l_2^+,g_0,m^2\}
\end{eqnarray*}
The enlarged operators must satisfy two conditions:\\
1)~They must form an algebra $[L_i,L_j]\sim{}L_k$;\\
2)~The operators which can't be regarded as constraints must
be zero or contain arbitrary parameters whose values will be
defined later from the condition of reproducing the correct
equations of motion.

In the case of higher spin fields in Minkowski space the algebra
of the operators is a Lie algebra
\begin{eqnarray}
[l_i,l_j]=f_{ij}^k\;l_k.
\label{aa}
\end{eqnarray}
In this case we can construct the additional parts of the operators
$l_i'$ which satisfy the same algebra (\ref{aa})
$[l_i',l_j']=f_{ij}^k\;l_k'$ using the method described in
\cite{AddP} and since the initial operators $l_i$ commute with the
additional parts $l_j'$ we get that the enlarged operators satisfy
the same algebra $[L_i,L_j]=f_{ij}^k\;L_k$ (\ref{aa}). After this
the BRST-BFV operators $Q'$ can be constructed in the usual way
(\ref{eq1}).

Now one need to define the arbitrary parameters. As explained in
\cite{0505092} we should assume that the state vectors
$|\Psi\rangle$ in the extended Fock space, including the ghost
fields, must be independent of the ghosts corresponding to the
Hermitian operators which are not constraints. Let us denote these
ghost as $\eta_G$ and $\eta_M$ corresponding to the extended
operators $G_0=g_0+g_0'$ and $M^2=m^2+m^{\prime2}$ respectively.

Let us extract the dependence of the BRST-BFV operator on the ghosts
$\eta_G$, $\mathcal{P}_G$, $\eta_M$, $\mathcal{P}_M$
\begin{eqnarray}
Q'&=&Q
+\eta_G(\sigma+h)+\eta_M (m^2+m^{\prime2})
-\eta_2^+\eta_2{\cal{}P}_G+\eta_1^+\eta_1 \mathcal{P}_M,
\label{Q'}
\end{eqnarray}
where $\sigma+h=g_0+g_0'+ \mbox{\it{}ghost fields}$, with $h$ and
$m^{\prime2}$ being the arbitrary parameters to be defined. After
this the equation on the physical states in the BRST-BFV approach
$Q'|\Psi\rangle=0$ yields three equations
\begin{eqnarray}
&&
Q|\Psi\rangle=0,
\qquad
(\sigma+h)|\Psi\rangle=0,
\qquad
(m^2+m^{\prime2})|\Psi\rangle=0
.
\label{eigenv}
\end{eqnarray}
From the last two equations in (\ref{eigenv}) we find the possible
values of $h$ and $m^{\prime2}$ whereas the first equation is
equation on the physical state. This equation on the physical state
can be obtained from the Lagrangian
\begin{eqnarray}
-\mathcal{L}=\int\!\! d\eta_0\; \langle\Psi|KQ|\Psi\rangle
.
\label{Qaction}
\end{eqnarray}
In eq. (\ref{Qaction}) above the standard scalar product in the
Fock space is used and $K$ is a specific invertible operator
providing the reality of the Lagrangian (see \cite{0505092}
for more details). The latter acts as the unit operator in the
entire Fock space, but for the sector controlled by the
auxiliary creation and annihilation operators used at
constructing the additional parts.

\section{Lagrangian construction for the fermionic fields}\label{sec:f}

The Lagrangian construction for the fermionic higher spin
theories have two specific differences compared to the bosonic ones
and demands some comments.

One of the specific features consists in that
we have
the fermionic operators in the algebra of constraints and
corresponding them the bosonic ghosts.
We can write these ghosts in any power in the Fock space
states.
As a result the resulting theory will be a gauge theory where the
order of reducibility grows with the spin of the field
(see \cite{0603212} for further details).

Another specific features is that unlike the bosonic case, in the
fermionic theory we must obtain Lagrangian which is linear in
derivatives. But if we try to construct Lagrangian similar to the
bosonic case (\ref{Qaction}) we obtain Lagrangian which will be the second order in
derivatives. To overcome this problem one first partially fixes the
gauge and partially solves some field equations. Then the obtained
equations are still Lagrangian and thus we can derive  the correct
Lagrangian (see \cite{0603212} for further details).

Using this method, the Lagrangians for the massive fermionic higher
spin fields have been obtained \cite{0603212}.

\section{Lagrangian construction for the fields in AdS}\label{sec:AdS}

The main difference of the Lagrangian construction in AdS space
is that  the algebra generated by the constraints is nonlinear,
but it has a special structure.
The structure of the algebra looks like \cite{0608005}
\begin{eqnarray}
[\,l_i,l_j]&=&
f_{ij}^kl_k+f_{ij}^{km}l_kl_m,
\label{alg-ini}
\end{eqnarray}
where $f_{ij}^k$, $f_{ij}^{km}$ are constants.
The constants $f_{ij}^{km}$ are proportional to the scalar
curvature and disappear in the flat limit.

We describe the method of finding the enlarged expressions for the
operators of the algebra (\ref{alg-ini}) \cite{0608005}, (see also
\cite{BL}). First, we enlarge the representation space by
introducing the additional creation and annihilation operators and
construct new operators of the algebra $l_i\to{}L_i=l_i+l_i'$,
where $l_i'$ is the part of the operator which depends on the new
creation and annihilation operators only (and constants of the
theory like the mass $m$ and the curvature).

Then
we
demand that the new operators $L_i$
 are in involution relations
\begin{eqnarray}
[\,L_i,L_j]&\sim& L_k.
\label{invL}
\end{eqnarray}
Since $[\,l_i,l_j']=0$ we have
\begin{eqnarray*}
[\,L_i,L_j]=
[\,l_i,l_j]+[\,l_i',l_j']
&=&
f_{ij}^kL_k
-(f_{ij}^{km}+f_{ij}^{mk})l_m'L_k+f_{ij}^{km}L_kL_m
\\
&&{}
-f_{ij}^kl_k'+f_{ij}^{km}l_m'l_k'
+[\,l_i',l_j']
.
\end{eqnarray*}
Then in order to provide (\ref{invL}) the last three terms must
be canceled. Thus we find the algebra of the
additional parts
\begin{eqnarray}
[\,l_i',l_j']
&=&
f_{ij}^kl_k'-f_{ij}^{km}l_m'l_k'
\label{addal}
\end{eqnarray}
and also we find the deformed algebra for
the enlarged operators
\begin{eqnarray}
[\,L_i,L_j]&=&
f_{ij}^kL_k
-(f_{ij}^{km}+f_{ij}^{mk})l_m'L_k+f_{ij}^{km}L_kL_m
.
\label{alg-enl}
\end{eqnarray}
We see that the algebra (\ref{alg-enl}) of the enlarged
operators $L_i$ is changed in comparison with the algebra
(\ref{alg-ini}) of the initial operators $l_i$.

There exists the method \cite{AddP} which allows us to construct explicit
expressions for the additional parts on the base of their
algebra (\ref{addal}).
Thus the problem of constructing of the additional parts for the
nonlinear algebra (\ref{alg-ini}) can be solved.
Let us remind that
the additional parts corresponding to operators which are not
constraints
must linearly contain arbitrary
parameters (whose values will be defined later from the condition
of reproducing the correct equations of motion) and therefore
the trivial solution is not allowed.

Next we discuss the aspects of constructing the BRST-BFV operator
caused by the nonlinearity of the operator algebra using the massive
bosonic higher spin fields in AdS space \cite{0608005}, \cite{BL} as
an example. The construction of BRST-BFV operator is based on
following general principles:
\\
{\bf 1.} The BRST-BFV operator $Q'$ is Hermitian, $Q^{\prime+}=Q'$,
and nilpotent, $Q^{\prime2}=0$.
\\
{\bf 2.} The BRST-BFV operator $Q'$ is built using a set of first
class constraints. In the case under consideration the operators
$\tilde{L}_{0}$, $L_{1}$, $L_{1}^{+}$, $L_{2}$, $L_{2}^{+}$, $G_{0}$
are used as a set of such constraints.
\\
{\bf 3.} The BRST-BFV operator $Q'$ satisfies the special initial
condition
\begin{eqnarray*}
Q'\Big|_{{\cal{}P}=0}
&=&
\eta_0\tilde{L}_0+\eta_1^+L_1+\eta_1L_1^+
+\eta_2^+L_2+\eta_2L_2^++\eta_{G}G_0
.
\label{Q-ini}
\end{eqnarray*}

Straightforward calculation of the commutators
allows us to find the algebra of the enlarged operators.
In particular for the bosonic fields in AdS space we get the following
commutation relations
\cite{0608005}
\begin{eqnarray}
[L_1,\tilde{L}_0]&=&
(\gamma-\beta) r L_1
+4\beta rL_1^+L_2
-4\beta rl_1^{\prime+}L_2
-4\beta rl_2'L_1^+
\nonumber
\\
&&{}
+2\beta r G_0L_1
-2\beta rl_1'G_0
-2\beta rg_0'L_1
,
\label{L1L0}
\\
{}
[\tilde{L}_0,L_1^+]
&=&
(\gamma-\beta) r L_1^+
+4\beta rL_2^+L_1
-4\beta rl_2^{\prime+}L_1
-4\beta rl_1'L_2^+
\nonumber
\\
&&{}
+2\beta r L_1^+G_0
-2\beta r l_1^{\prime+}G_0
-2\beta r g_0'L_1^+
,
\label{L0L1+}
\\
{}
[L_1,L_1^+]
&=&
\tilde{L}_0-\gamma rG_0
+4(2-\beta)r(l_2^{\prime+}L_2+l_2'L_2^+)
\nonumber
\\
&&{}
-2(2-\beta)rg_0'G_0
+(2-\beta)r(G_0^2-2G_0-4L_2^+L_2)
.
\label{L1L1+}
\end{eqnarray}
All possible ways to order the operators in the right hand sides of
(\ref{L1L0})--(\ref{L1L1+}) can be described in terms of arbitrary
real parameters $\xi_1$, $\xi_2$, $\xi_3$, $\xi_4$, $\xi_5$. The
arbitrariness in the BRST-BFV operator caused by the parameter
$\xi_i$ is resulted in arbitrariness of introducing the auxiliary
fields in the Lagrangians and hence does not affect the dynamics of
the basic field (see \cite{0608005} for the details). After that,
the construction of the Lagrangians for the fields in AdS space goes
the practically the same way as for fields in Minkowsky space.

Using this method, the  Lagrangians for the bosonic \cite{0608005}
and for fermionic \cite{0703049} massive higher spin fields in AdS
space have been constructed.

\section{Fields corresponding to an arbitrary Young tableau}\label{sec:Young}

Now we consider the Lagrangian construction for the fields
corresponding to non sqaure Young tableau using a Young tableau with
2 rows ($s_1 \ge s_2$)
\begin{eqnarray}
\Phi_{\mu_1 \cdots \mu_{s_1},\,\nu_1 \cdots \nu_{s_2}} (x)
&\longleftrightarrow&
\begin{tabular}{|l|l|l|l|ll}
  \hline
  $\mu_1$ & $\mu_2$ & $\cdots$ & $\cdots$
  &\multicolumn{2}{l|}{$\cdots$ \vline $\;\;\mu_{s_1}$}\\
  \hline
  $\nu_1$ & $\nu_2$ & $\cdots$ & $\nu_{s_2}$ &  & \\
  \cline{1-4}
\end{tabular}
\label{basic}
.
\end{eqnarray}
The tensor field is
symmetric with respect to permutation of each type of the
indices\footnote{The indices inside round brackets are to be
symmetrized.}
\begin{math}
\Phi_{\mu_1 \cdots \mu_{s_1},\,\nu_1 \cdots \nu_{s_2}} (x)
=
\Phi_{(\mu_1 \cdots \mu_{s_1}),\,(\nu_1 \cdots \nu_{s_2})} (x)
\end{math}
and in addition must satisfy the following equations
\begin{eqnarray}
&&
( \partial^2 + m^2 )
\Phi_{\mu_1\cdots\mu_{s_1},\,\nu_1\cdots\nu_{s_2}} ( x ) = 0
,
\label{KG}
\\
&&
\partial^{\mu_1}
\Phi_{\mu_1\cdots\mu_{s_1},\,\nu_1\cdots\nu_{s_2}} ( x ) =
\partial^{\nu_1} \Phi_{\mu_1
\cdots\mu_{s_1},\,\nu_1\cdots\nu_{s_2}} ( x ) = 0
\label{transversal}
,
\\
&&
\eta^{\mu_1\mu_2}
\Phi_{\mu_1 \mu_2\cdots\mu_{s_1},\,\nu_1\cdots\nu_{s_2}}
=
\eta^{\nu_1\nu_2}
\Phi_{\mu_1\cdots\mu_{s_1},\,\nu_1\nu_2\cdots\nu_{s_2}}
=
\eta^{\mu_1\nu_2}
\Phi_{\mu_1\cdots\mu_{s_1},\,\nu_1\cdots\nu_{s_2}}
=0
\label{traceless}
,
\\
&&
\Phi_{(\mu_1 \cdots \mu_{s_1},\,\nu_1) \cdots \nu_{s_2}} (x)
=0.
\label{const4}
\end{eqnarray}
Then we define Fock space generated by creation and
annihilation operators
\begin{eqnarray}
[a_i^{\mu}, a_j^{+\nu} ] = - \eta^{\mu \nu} \delta_{ij},
\qquad
\eta^{\mu \nu} = {diag} ( +,-, -, \cdots, - )
\qquad
i,j=1,2.
\label{[]}
\end{eqnarray}
The number of pairs of creation and annihilation operators one
should introduce is determined by the number of rows in the
Young tableau corresponding to the symmetry of the tensor field.
Thus we introduce two pairs of such operators.
An arbitrary state vector in this Fock space has the form
\begin{eqnarray}
|\Phi\rangle
&=&
\sum_{s_1=0}^{\infty} \sum_{s_2=0}^{\infty}
\Phi_{\mu_1\cdots\mu_{s_1},\,\nu_1\cdots\nu_{s_2}}(x)\;
a_1^{+\mu_1} \cdots a_1^{+\mu_{s_1}} a_2^{+\nu_1}\cdots
a_2^{+\nu_{s_2}}
|0\rangle
\label{Phi>}
.
\end{eqnarray}
To get equations
(\ref{KG})--(\ref{const4}) on the coefficient functions
we introduce the following operators
\begin{align}
&
l_0 = - p^\mu p_\mu + m^2,
&&
l_i = a_i^{\mu} p_{\mu},
&&
l_{{ij}} = \frac{1}{2} a_i^{\mu} a_{j \mu}
&&
g_{12} = - a_1^{+\mu} a_{2 \mu}
\label{g12}
\end{align}
where $p_\mu=-i\partial_\mu$. One can check that restrictions
(\ref{KG})--(\ref{const4}) are equivalent to
\begin{align}
&
l_0|\Phi\rangle=0\,,
&&
l_i|\Phi\rangle=0\,,
&&
l_{ij} | \Phi\rangle=0\,,
&&
g_{12}|\Phi\rangle=0
\label{constraints}
\end{align}
respectively.

Now we can generalize this construction to the fields corresponding
to $k$-row Young tableau. For this purpose one should introduce Fock
space generated by $k$ pairs of creation and annihilation operators
(\ref{[]}), where $i,j=1,2,\ldots,k$, and then introduce
operators\footnote{Operator $g_{12}$ is generalized to operator
$g_{ij}=-a_i^{+\mu} a_{j\mu}$ where $i>j$.} (\ref{g12}), but now
with $i,j=1,2,\ldots,k$. After this the Lagrangian construction can
be carried out as usual \cite{0707.2181}. Using this method
Lagrangians for the massive bosonic field corresponding to 2-rows
Young tableau was constructed in \cite{0707.2181}.

\section{Summary}\label{Summary}

In this paper we have briefly considered the basic principles of
gauge invariant Lagrangian construction for massive higher spin
fields. This method can be applied to any free higher spin field
model in Minkowski and AdS spaces. It is interesting to point out
that the Lagrangians obtained possess a reducible gauge invariance
and for the fermionic fields the order of reducibility grows with
value of the spin. Recent applications of BRST-BFV approach to
interaction higher spin theories are discussed in \cite{INT}.

\vspace{1em}
{\bf Acknowledgements.}
The work of I.L.B
and V.A.K was partially supported by the INTAS grant, project
INTAS-05-7928, the RFBR grant, project No.\ 06-02-16346 and grant
for LRSS, project No.\ 4489.2006.2. Work of I.L.B was supported in
part by the DFG grant, project No.\ 436 RUS 113/669/0-3 and joint
RFBR-DFG grant, project No.\ 06-02-04012. 


\end{document}